\documentclass[prd,superscriptaddress,onecolumn,showpacs,nofootinbib]{revtex4}
\usepackage{amssymb}
\usepackage{graphicx}
\usepackage{dcolumn}
\usepackage{bm}

\usepackage{graphicx}
\usepackage{amsmath}
\usepackage[usenames]{color}
\usepackage{epsfig}
\def\bea{\begin{eqnarray}}
\def\eea{\end{eqnarray}}

\newcommand{\na}{\nabla}
\def\beq{\begin{equation}}
\def\eeq{\end{equation}}

\def\pa{\partial}

\def\na{\nabla}

\def\M{\mu}

\def\bth{\mbox{\boldmath $\theta$}}
\begin{document}

\title{Dark aspects of massive spinor electrodynamics }
\author{Edward J. Kim}\email{jkim575@illinois.edu}
\affiliation{Department of Physics and Institute of Basic Science,
 Sungkyunkwan University, Suwon 440-746, Korea}
\affiliation{Department of Physics, University of Illinois, Urbana, IL 61801 USA}
\author{Seyen Kouwn}\email{seyen@ewha.ac.kr}
\affiliation{Department of Physics and Institute of Basic Science,
 Sungkyunkwan University, Suwon 440-746, Korea}
\affiliation{Institute for the Early Universe, Ewha Womans University, Seoul 120-750, South Korea}
\author{ Phillial Oh}\email{ploh@skku.edu}
\affiliation{Department of Physics and Institute of Basic Science,
 Sungkyunkwan University, Suwon 440-746, Korea}
\author{Chan-Gyung Park}\email{parkc@jbnu.ac.kr}
\affiliation{Division of Science Education and Institute of Fusion Science,
Chonbuk National University, Jeonju 561-756, Korea}

\begin{abstract}
 We investigate the cosmology of massive spinor electrodynamics when torsion is non-vanishing. A non-minimal interaction is introduced between the torsion and the vector field and the coupling constant between them plays an important role in subsequential cosmology. It is shown that the mass of the vector field and torsion conspire to generate dark energy and pressureless dark matter, and for generic values of the coupling constant, the theory effectively provides  an interacting  model between them  with an additional
energy density  of the form $\sim 1/a^6.$ The evolution equations mimic $\Lambda$CDM
behavior up to $1/a^3$ term and the additional term represents a deviation from $\Lambda$CDM. We  show that the deviation is compatible with the observational data, if it is very small. We find that the non-minimal
interaction is responsible for generating an effective cosmological constant  which is directly proportional to the mass squared of the vector field
and     the mass of the  photon within its current observational limit could be the source of the dark energy.
\end{abstract}
\pacs{98.80.-k, 95.36.+x}
\maketitle
%

\section{Introduction}
One of the most intriguing discovery of modern cosmology is the acceleration of
the Universe \cite{Riess:1998cb,Perlmutter:1998np}.
A    standard approach is to assume  that dark energy of repulsive nature is causing the current acceleration.
Many candidates  of the dark energy have been proposed \cite{Carroll:2003qq,Copeland:2006wr,Sahni:2006pa}, among which the cosmological constant is the most accepted one.
 Along with yet another unidentified constituent of the Universe called dark matter, they
compose standard cosmological model, $\Lambda$CDM \cite{Komatsu:2008hk}.
Even though the extreme fine-tuning of the cosmological constant \cite{Weinberg:1988cp} has been an unsatisfactory theoretical feature of the model, and many alternatives  to explain the smallness of the cosmological constant as the source of dark energy have been proposed,
  it is remarkable that so far the observable Universe can be well addressed with the $\Lambda$CDM model.

In this paper, we investigate cosmology of massive spinor electrodynamics
with torsion. The theory consists of massive vector field interacting with the Dirac spinor in the Einstein-Cartan space-time,
and we find  that both of these fields  and torsion contribute to the energy densities of
dark energy and pressureless dark matter at late times.

Spinor electrodynamics in curved space-time is an old subject and there
are extensive amounts of literature on the subject, describing both classical and quantum aspect \cite{toms}.
But as far as cosmology is concerned, vector sector and spinor sector were considered separately so far.
On the spinor side,
it has been known for some time that  spinor fields could play some  important roles in the evolution of the Universe \cite{ferm,picon,riba,saha}.
  For example, in Ref. \cite{picon}  it is shown that a spinor field can accommodate any desired behavior of its energy density if
an appropriate  self-interaction of the spinor field is introduced.
For the Dirac spinor mass term, the energy density shows a cosmological behavior exactly like a pressureless dark matter term.
In Ref. \cite{riba} it is investigated whether fermionic sources could be responsible for accelerated periods during
the evolution of a universe after a matter field would guide for the decelerated period.   In Ref. \cite{saha} the authors have shown
that it is possible to simulate perfect fluid and dark energy by means of a nonlinear spinor field.

On the other hand, the cosmology with  vector fields received more attention  recently and various models have been proposed to account for dark energy
\cite{ford}.
Nonlinear electromagnetism in cosmology was considered in Ref. \cite{Novello:2003kh},
where it was shown that the addition of a non-linear term to the Lagrangian of the electromagnetic field yields a fluid with an asymptotically super-negative equation of state, causing an accelerated
expansion of the universe.
Universe filled with a massive vector field non-minimally coupled to gravitation
was proposed, and  the cosmology yields a dark energy component which is proportional to the mass of the vector particle \cite{boehmer}.
In Ref. \cite{Koivisto:2008xf}, several (non)-minimally coupled vector field models with a potential for the vector field  were considered and
some model mimics  $\Lambda$CDM expansion at late times.
In particular,
 the possibility of understanding dark energy from the standard electromagnetic field, without the need of introducing new physics was explored recently \cite{maroto} and it was shown that the presence of a temporal electromagnetic field on cosmological
scales generates an effective cosmological constant which can account for the accelerated
expansion of the universe.

In this paper we attempt investigations on the  cosmological consequences of the massive spinor electrodynamics where both  spinor and vector sectors come into play, but they are non-minimally coupled to gravitation.
One could naively expect that the pressureless dark matter term \cite{picon} coming from the spinor  sector and the effective cosmological constant  \cite{maroto} coming from the massless vector sector could combine
to yield a dark energy model whose cosmological evolution mimics that of the standard $\Lambda$CDM model.
 However, one finds out immediately  that this anticipation is not met by an explicit check
of the equations of motion, because the theory has an interaction between
the spinor and the  vector  field. It turns out that such a difficulty can be overcome by introducing a massive vector field and torsion \cite{hamm} component along with a specific non-minimal interaction between them into the theory. They  can intervene
between the two sectors and alleviate  the difficulties  coming from the interaction.
Consequently, we find that cosmology of massive spinor electrodynamics provides a dark energy model where the dark energy and dark matter are interacting with each other through the electromagnetic interaction.
It is found that the model allows  an asymptotic de Sitter acceleration, which is an attractor. When dark energy dominance has taken place at late times, the spinor provides dark matter density, whereas the massive vector field and torsion contribute to both dark energy and dark matter densities.
One of the interesting aspects as a consequence is that the dark energy density is directly proportional to the mass squared of the vector fields.
Another distinctive feature coming from the cosmology of massive spinor
 electrodynamics is that  the dynamics allows  a very small fraction of the energy density evolving as $\sim 1/a^6$ at late times.
We check whether the presence of this energy density  is consistent with the observational data
by giving a detailed analysis of how much of such a deviation from $\Lambda$CDM  can be allowed.

The paper is organized as follows: In Sec. 2, we give a classical formalism of massive spinor electrodynamics in curved space-time with torsion.
In Sec. 3, we investigate the cosmology of massive spinor electrodynamics
and show that it can provide a field theoretical model of dark energy and dark matter.  We find de Sitter solution and  perform an asymptotic expansion to show  that an additional  energy density component of the form $\sim 1/a^6$
is present in this approach.
In Sec. 4,  we present data analysis in order to test whether  $\Lambda$CDM+
$(1/a^6)$ can be consistent with observations.
 Sec. 5 includes conclusion and discussions.

\section{Massive spinor electrodynamics with torsion}

We start with a brief summary of torsion. Let us consider the connection
\begin{eqnarray}
\Gamma^\rho_{\mu\nu}=\big\{
\begin{matrix}\rho \\
\mu\nu\ \\
\end{matrix}\big\}-K^{~~~\rho}_{\mu\nu},\label{torsionconnection}
\end{eqnarray}
where $\big\{
\begin{matrix}\rho \\
\mu\nu \\
\end{matrix}\big\}$ is the Christoffel connection  and the contortion $K^{~~~\rho}_{\mu\nu}$ is given by the torsion tensor
$S^{~~~\rho}_{\mu\nu}=\Gamma^\rho_{[\mu\nu]}$ via
\begin{eqnarray}
K^{~~~\rho}_{\mu\nu}= -\left(S^{~~~\rho}_{\mu\nu}+S^{\rho}_{~\mu\nu}+S^{\rho}_{~\nu\mu}\right).\label{contorsin}
\end{eqnarray}
The above connection (\ref{torsionconnection}) satisfies the metricity condition,
$\nabla_\mu g^{\nu\rho}=0$.
It is important to note that $K_{\alpha\beta\gamma}$ is anti-symmetric for last two indices
 and $S_{\alpha\beta\gamma}$ anti-symmetric for first two
 indices.
 Now, we can generally decompose the contortion tensor (\ref{contorsin}) into
 a traceless part and trace \cite{sa}:
\begin{equation}
{K_{\mu\nu}}^{\rho} = {\tilde{K}_{\mu\nu}}^{~~~\rho} -
\frac{2}{3}\left( \delta_{\mu}^{\rho} S_\nu  -
g_{\mu\nu} S^\rho \right), \label{decompos}
\end{equation}
where $\tilde{K}_{\mu\nu}^{\ \ \mu}=0$
 and $S_\nu$ is
the trace of the torsion tensor, $S_\nu = S^{\ \
\mu}_{\mu\nu}$.
Making use of the
connection ${\Gamma}_{\mu\nu}^\rho$ with
(\ref{decompos}),
  we can write curvature scalar as follows
\begin{eqnarray}
R = R(\{\}) - 4\stackrel{\{\}}{\nabla}_\mu S^\mu -\frac{8}{3}S_\mu S^\mu -\tilde{K}_{\nu\rho\alpha} \tilde{K}^{\alpha\nu\rho},\label{r2}
\end{eqnarray}
with the obvious notation that quantities with $\{\}$ are those which are constructed with  Christoffel connection only.

Next, let us consider an action  given by (in units of  $\sqrt{1/8\pi G}\equiv M_p=1$)
\begin{eqnarray}
S^\prime = \int d^4x~ \sqrt{-g} \biggr[{ R \over 2} - {1\over4}F_{\mu\nu}F^{\mu\nu}-\frac{1}{2}m^2A_\mu A^\mu
\biggr] \,+\int d^4 x ~\sqrt{-g}~{\cal L}_{D}\, , \label{action}
\end{eqnarray}
\begin{eqnarray}
{\cal L}_{D}=-\frac{i}{2} \left[(\nabla_\mu\bar\psi)\gamma^\mu\psi-\bar\psi\gamma^\mu \nabla_\mu\psi\right] -m_f \bar\psi\psi+A_\mu \bar\psi\gamma^\mu\psi.\label{dirac}
\end{eqnarray}
The above action describes a massive vector field which is known as Proca field interacting with Dirac fermion in curved spacetime.
For the field strength, we can choose the gauge invariant non-minimal coupling prescription \cite{hamm}
\begin{eqnarray}
F_{\mu\nu}=\partial_\mu A_\nu-\partial_\nu A_\mu,\label{fss}
\end{eqnarray}
or  the minimal coupling, $\tilde F_{\mu\nu}=\nabla_\mu A_\nu-\nabla_\nu A_\mu=F_{\mu\nu}-2S^{~~~\rho}_{\mu\nu}A_\rho$, which is not gauge invariant.
For the case of Proca field, the  term $\sim S^{~~~\rho}_{\mu\nu}A_\rho $ is harmless, since there is no gauge invariance. However,   the minimal coupling 
encounters  problems in the massless limit, because with this term present the field strength is no longer gauge invariant in that limit. Therefore, we adopt non-minimal expression (\ref{fss}) which allows a smooth massless limit{\footnote{The
non-minimal prescription adopted here
might confront conceptual difficulty
because vector fields are decoupled
from torsion, although they are spinning particles. The situation could be improved by adding an explicit interaction term
between the vector field and torsion \cite{gasp}. However, such interactions do not play any role, because $F_{\mu\nu}=0$, for the background cosmological configuration in our case. It is also pointed out that as far as cosmology is concerned, the two approaches do not  make any difference. }}.
The  limit involves  another important modification of the action
(\ref{action}) itself.     In the flat spacetime, the Proca theory is quantum mechanically consistent  as long as it couples to a conserved current, because the condition  $\partial_\mu A^\mu=
1/m^2 \partial_\mu J^\mu= 0$
 eliminates the unwanted ghost state of the vector field. However, the model has problem when massless  limit is considered,  that is,  a singularity is encountered  in the  limit \cite{zuber}. This is reflected in the fact that the limiting Lagrangian $-\frac{1}{4}F^2$ with gauge symmetry is unsuitable without a gauge prescription and massless spinor electrodynamics is not recovered in that limit.
To recover the internal symmetry structure smoothly in the limit,   an auxiliary condition which behaves
like a gauge fixing term in the massless limit $(\sim  (\partial\cdot A)^2 )$ is imposed. In curved background, we introduce
\begin{eqnarray}
S_{gf} =-\frac{1}{2\alpha}\int d^{4}x\sqrt{-g}\left
[\left(\stackrel{\{\}}{\nabla}_{\mu}A^{\mu}\right)^{2}+4\xi S_{\mu}A^{\mu}
\stackrel{\{\}}{\nabla}_{\nu}A^{\nu}+4 \xi^2
\left(S_{\mu}A^{\mu}\right)^{2}\right], ~~~S=S^\prime +S_{gf}.
\label{gafi}
 \end{eqnarray}
A couple of comments are in order regarding the above auxiliary condition (or gauge fixing term in the massless limit) for the massive vector field. In flat space with vanishing torsion, this term reduces to $-\frac{1}{2\alpha} (\partial\cdot A)^2$ which guarantees a smooth massless limit $m\rightarrow 0$ for the propagator, and massless spinor electrodynamics is recovered in the limit.
$\alpha\rightarrow \infty$ gives the ordinary
 Proca  theory.
 When $\xi=1,$ three terms of (\ref{gafi}) combine into a single covariant derivative with torsion connection $\sim (\nabla_\mu A^\mu)^2$ which corresponds to a minimal extension to spacetime with torsion. $\xi\neq 1$ can be regarded as a non-minimal coupling  between the vector field and torsion.
Even though  $\xi=1$  corresponds to extending the gauge fixing term straightforwardly to curved space with torsion,
there is no a priori reason to favor this choice and  we consider general values of $\xi$ which turns out to be important to account for  dark energy. Since $\xi\rightarrow -\xi$ can always be compensated by $S_\mu\rightarrow -S_\mu$, we will assume $\xi\geq 0$ without loss of generality.
If we rescale $S_\mu\rightarrow \xi^{-1} S_\mu,$
$\xi$-dependence disappears in (\ref{gafi}), but reappears in (\ref{r2}).
We stick to the expression of (\ref{gafi}).

  For the Dirac part,  we use the gamma matrices given by
\bea
\gamma^0= \left(
           \begin{array}{cc}
             1 & 0 \\
             0 & -1 \\
           \end{array}
         \right),~~\gamma^i=\left(
                              \begin{array}{cc}
                                0 & \sigma_i \\
                                \ -\sigma_i & 0 \\
                              \end{array}
                              \right)
                           \eea
 with
         $\{\gamma^a, \gamma^b\}=-2\eta^{a b},~~ \eta^{a b}= \text{diag}.(-1,1,1,1).$
       The covariant derivative of the spinor and
its dual are given by
\beq \na_{\M}\psi \equiv \pa_{\M}\psi-\Gamma_{\M}{\psi}
 \eeq
  and
\beq \na_{\M}{\bar\psi}
\equiv\pa_{\M}\bar{\psi}+\bar{\psi}\Gamma_{\M}, \eeq
where $\Gamma_{\M}$ is the connection on the spinor given by
\bea
\Gamma_\mu = \frac{1}{4} \omega_{\mu}^{~ab} \gamma_a \gamma_b.\label{spinorconnection}
\eea
The spin connection is given by
\bea
\omega_{\mu}^{~ab} = e^a_\nu\left( \partial_\mu e^{b\nu} + \Gamma^\nu_{\mu\rho}  e^{b\rho}\right),\label{spinconnection}
\eea
where  the tetrad is defined by
\bea
g_{\mu\nu}=e^a_\mu e^b_\nu \eta_{ab}.
\eea
One can check that the spin connection  satisfies $\omega_{\mu~b}^{~a}=-\omega_{\mu b}^{~~~a}$.
 It is useful to decompose the spin connection (\ref{spinconnection}) into two parts,
 \bea
 \omega_{\mu}^{~ab}={\omega}_{\mu}^{~ab}(\{\})-e_\nu^a e^{b\rho}K^{~~~\nu}_{\mu\rho}.
 \eea
Then, one can show that Eq. (\ref{dirac}) becomes \cite{hehl1}
\bea
{\cal L}_{D}={\cal L}_{D}(\{\})+\frac{i}{4}\bar\psi\gamma^{[\rho}\gamma^{\nu}\gamma^{\mu]}
\psi K_{\mu\nu\rho}.
\eea
Hence, the Dirac spinor interacts only with the totally anti-symmetric components of the torsion.

The equations of motions for $g_{\mu\nu}$ are given by
\begin{eqnarray}
G_{\mu\nu}(\{\}) =  T_{\mu\nu}(T,A)+T_{\mu\nu}(D)  \,,\label{einstein}
\end{eqnarray}
where
\begin{eqnarray}
T_{\mu\nu}(T,A) & =&\frac{8}{3}\left[S_{\mu}S_{\nu}-\frac{1}{2}g_{\mu\nu}S_{\alpha}S^{\alpha}\right]
 +\left[{{\tilde K}_{\alpha(\mu}}^{~~~~\beta} {\tilde K_{\nu)\beta}}^{~~~~\alpha} - \frac{1}{2}g_{\mu\nu}\tilde K_{\alpha\beta\gamma}\tilde K^{\alpha\beta\gamma}\right]\\
  && +\left[F_{\mu\alpha}F_{\nu}{}^{\alpha}-\frac{1}{4}g_{\mu\nu}F_{\alpha\beta}F^{\alpha\beta}
  \right]+m^2\left[A_\mu A_\nu-\frac{1}{2}g_{\mu\nu} A_\alpha A^\alpha \right]\\
 && +\frac{g_{\mu\nu}}{\alpha}\left[\frac{1}{2}\left(\stackrel{\{\}}{\nabla}_{\mu}A^{\mu}\right)^{2}
 +A^{\nu}\stackrel{\{\}}{\nabla}_{\nu}\left(\stackrel{\{\}}{\nabla}_{\alpha}A^{\alpha}+2\xi S_\alpha A^\alpha\right)-
 2\xi^2 \left(S_{\alpha}A^{\alpha}\right)^{2}\right]\nonumber \\
 && +\frac{1}{\alpha}\left[-2A_{(\mu}\stackrel{\{\}}{\nabla}_{\nu)}
 \left(\stackrel{\{\}}{\nabla}_{\alpha}A^{\alpha}+2\xi A_\alpha S^\alpha\right)+
 4\xi S_{(\mu}A_{\nu)}\stackrel{\{\}}{\nabla}_{\alpha}A^{\alpha}
 +8\xi^2 S_{(\mu}A_{\nu)}\left(S_{\alpha}A^{\alpha}\right)\right],\nonumber
\end{eqnarray}
and
\bea
T_{\mu\nu}(D)=\frac{i}{4}\left[(D_\mu\bar\psi)\gamma_\nu\psi
+(D_\nu\bar\psi)\gamma_\mu\psi-
\bar\psi\gamma_\mu D_\nu\psi
-\bar\psi\gamma_\nu D_\mu\psi
\right]
+g_{\mu\nu}{\cal L}_{D}(\{\}),
\eea
where ${\cal L}_{D}$ is given in (\ref{dirac}) and we introduced $D_\mu=\stackrel{\{\}}\nabla_\mu
-iA_\mu$.
Equation of motions for $A_{\mu}$ is given by
\begin{equation}
0=\frac{1}{\sqrt{-g}}\partial_{\nu}\left(\sqrt{-g}F^{\mu\nu}\right)+m^2A^\mu-
\frac{1}{\alpha}\left[{\stackrel{\{\}}\nabla} ~^\mu
\left(\stackrel{\{\}}{\nabla}_{\nu}A^{\nu}\right)-2\xi S^{\mu}\stackrel{\{\}}{\nabla}_{\nu}
A^{\nu}+2\xi{\stackrel{\{\}}{\nabla}}~^{\mu}
\left(S_{\nu}A^{\nu}\right)-4\xi^2S^{\mu}S_{\nu}A^{\nu}\right]-\bar\psi\gamma^\mu\psi,\label{maxwell}
\end{equation}
and the equation coming from variation of the torsion $S_\mu$ is
\bea
S^\mu =-\frac{3\xi}{4\alpha}\left[A^{\mu}\stackrel{\{\}}{\nabla}_{\nu}A^\nu
+2\xi A^{\mu}S_{\nu}A^{\nu}\right],\label{torsioneq}
\eea
and for the totally antisymmetric component $K_{[\mu\nu\rho]}$, we have
\bea
K_{[\mu\nu\rho]}=\frac{i}{4}\bar\psi\gamma^{[\rho}\gamma^{\nu}\gamma^{\mu]}
\psi.\label{antieq}
\eea
The Dirac equation is given by
\bea
\left(i\gamma^\mu\stackrel{\{\}}{\nabla}_\mu-m_f+\gamma^\mu A_\mu\right)\psi=
-\frac{i}{4}\gamma^{[\rho}\gamma^{\nu}\gamma^{\mu]}
\psi K_{\mu\nu\rho}.\label{diracm}
\eea

\section{FRW Cosmology }
To discuss Friedman cosmology in the flat Robertson-Walker space-time,
consider a metric of the form
\begin{eqnarray}
ds^2=-dt^2+ a^2(t) dx^i dx^i,\label{RW}
\end{eqnarray}
where $a(t)$ is the scale factor
of our universe.
Isotropy and homogeneity permit only the following non-vanishing components
of  torsion \cite{tsam}
\begin{eqnarray}
S^{~~~1}_{10}=S^{~~~2}_{20}=S^{~~~3}_{30}=h(t)/2,~\tilde K_{[ijk]}=\epsilon_{ijk}k(t)/6. \label{torsion}
\end{eqnarray}
 Here $h\left(t\right)$ and $k(t)$ are unknown functions of time whose dynamics are governed by the   equations of motion.
Before proceeding, we notice that the second ansatz of Eq. (\ref{torsion})  and (\ref{antieq}) yields
\bea
k(t)=-\frac{1}{4}\bar\psi \gamma_0\gamma^5\psi,
\eea
which vanishes for spinor configurations (see below) which respect spatial isotropy. Therefore,
the totally antisymmetric part of torsion is dropped from now on. Then, RW metric gives the following non-vanishing connection components
\bea
\big\{
\begin{matrix}0 \\
ij \\
\end{matrix}\big\} =a\dot a\delta_{ij},~~~
\big\{
\begin{matrix}i \\
j0 \\
\end{matrix}\big\}=\frac{\dot a}{a}\delta_{ij}.
\eea
and the tetrad with
\bea
e^a_\mu=(1, a, a, a),~~ e_a^\mu = (1, \frac{1}{a}, \frac{1}{a}, \frac{1}{a}),
\eea
which in turn produces the following  non-vanishing components of the spin connection (\ref{spinconnection}):
\bea
\omega_{x^i~0}^{~~j}(\{\})=\omega_{x^i~j}^{~~0}(\{\})=\dot a\delta_{ij}.
\eea
Then, we have from Eq. (\ref{spinorconnection})
\bea
\Gamma_0(\{\})=0,~~\Gamma_{x^i}(\{\})= \frac{\dot a}{2}\gamma_0\gamma_i.
\eea
For the gauge field, we choose a non-vanishing temporal gauge field  for isotropic configuration \cite{maroto}
\begin{eqnarray}
A_\mu = \left(  f(t),0,0,0 \right) \,.
\end{eqnarray}

Using these ingredients,  Eqs. (\ref{einstein}), (\ref{maxwell}), (\ref{torsioneq}), and (\ref{diracm}) become $(H=\dot a/a)$
\bea
3H^{2}&=&-\frac{1}{2\alpha}\left[\dot{f}+3\left
(H+\xi  h\right)f\right]^{2}+3h^{2}-\frac{m^2}{2}f^2-
f\bar\psi\gamma^0\psi+m_f\bar\psi\psi\label{energyeq}
\\
-3H^{2}-2\dot{H}&=&\frac{1}{2\alpha}\left[\dot{f}+3\left(H+\xi  h\right)f\right]^{2}+
3h^{2}-\frac{m^2}{2}f^2-f\bar\psi\gamma^0\psi\label{pressureeq}
\\
0&=&\partial_{t}\left(\dot{f}+3\left(H+\xi  h\right)f\right)-
3\xi h\left(\dot{f}+3\left(H+\xi  h\right)f\right)+\alpha m^2 f+\alpha \bar\psi\gamma^0\psi
\label{cosmomax}\\
\frac{2\alpha}{\xi}\frac{h}{f}&=&\dot{f}+3\left(H+\xi  h\right)f
\label{constrainttorsion}\\
0&=&\dot\psi+\frac{3}{2}H\psi+i\gamma^0 m_f \psi-i f\psi.\label{cosmodirac}
\eea
In deriving Eqs.  (\ref{energyeq}) and (\ref{pressureeq}), Eq. (\ref{constrainttorsion}) was used.

Let us discuss solutions of the evolution equations.  Eq. (\ref{cosmodirac}) gives
\bea
\bar\psi\psi=\frac{A}{a^3}, ~~\psi^\dagger\psi=\frac{B}{a^3},
\eea
where $A$ and $B$ are constants. They are related by $A=B(>0)$  for  spinor configuration
$\psi=(\psi_1, 0,0,0)$ and  by $A=-B (<0)$  for
$\psi=(0, 0,0,\psi_4)$  with constants  $A$ and $B$. Each respects isotropy with $\bar\psi \gamma^i\psi=0$. This solution is independent of the rest of the equations. We choose $A=B$ from here on, for which the fermionic mass term in Eq. (\ref{energyeq}) contributes positive energy density.
For the remaining four evolution equations, one can explicitly check that only three of them are independent for the three variables $a(t), ~f(t)$
and $h(t).$
 We also notice that the first term in the right hand side  of (\ref{energyeq})  corresponds to equation of state $\omega=-1$, middle three terms to $\omega=1$, and the last term to pressureless matter. However, one is not to be led to the conclusion that the first term is a constant and the combined middle three terms behaves $\sim 1/a^6$, because these are true only when each term is considered separately.

 We  first discuss an existence of an asymptotic  de Sitter phase
 which  is obtained by neglecting the decaying energy densities of spinor contributions.
 Assume $\alpha$ to be negative ($\alpha=-\vert\alpha\vert$) and  define
\bea \sqrt{\Lambda(t)}\equiv \sqrt{\frac{1}{2\vert\alpha\vert}}
\left[\dot{f}+3\left(H+\xi  h\right)f\right].\label{cosco}
\eea
Then, Eqs.   (\ref{energyeq}) and (\ref{pressureeq}) suggest that $\Lambda$ behaves as a cosmological constant ($\rho=-p$)
given by
\bea
\Lambda=\frac{\vert\alpha\vert}{3\xi^2}m^2, \label{coss}
\eea
and
 both $h$ and $f$ are constants being related by
\bea 3h_*^2=\frac{m^2}{2} f_*^2.\label{hfconstants}
\eea
 Eqs.   (\ref{cosmomax}) and (\ref{constrainttorsion}) produce a relation which is satisfied by $\Lambda$ given above when the spinor contributions are neglected. The remaining job is to check whether the Hubble parameter $H$ obtained from (\ref{constrainttorsion}) is consistent with (\ref{energyeq}). Equating both equations and using (\ref{hfconstants}), we have
\bea
h_*^4-\left(2-\frac{1}{\xi^2}\right)\frac{\alpha m^2}{9\xi^2}h_*^2 +\frac{\alpha^2 m^4}{81\xi^4}=0,
\eea
from which we have
\bea
h_*^2(\xi) =\frac{\vert\alpha \vert  m^2}{18\xi^4}\left[1-2\xi^2
\pm\sqrt{1-4\xi^2}\right],
~f_*(\xi) = {\sqrt{\vert\alpha\vert}\over \sqrt{6}\,\xi^2}\left(
1\pm\sqrt{1-4\xi^2}
\right).\label{solex}
\eea
Therefore, $\vert\xi\vert^2\leq 1/4$ and we have $f_*(\xi)\geq 0$.
Note that the cosmological constant in (\ref{coss}) is directly proportional to mass squared of the vector field and diverges when $\xi\rightarrow 0$. This is to be compared with the massless case \cite{maroto}
where $\Lambda$ is a priori undetermined constant except  its dependence on the gauge fixing parameter $\alpha$.

To proceed further, we first note that in the constant solution with $h=h_*$  and $f=f_*$
the spinor contribution and time dependence of each energy densities are neglected.
To account for these,  we  consider an expansion in terms of negative powers of the scale factor around the solution (\ref{solex})
in the dark energy dominance epoch,
\begin{eqnarray}
f=f_{*}(\xi)\left(1 +\frac{f^{(3)}}{a^3}+\frac{f^{(6)}}{a^6}+\cdots\right), ~~h=h_{*}(\xi)\left(1 +\frac{h^{(3)}}{a^3}+\frac{h^{(6)}}{a^6}+\cdots\right).
\label{expansion}
\end{eqnarray}
Then, we  adopt the following systematic method: (i)
Substitute the expressions (\ref{expansion}) and (\ref{constrainttorsion}) into (\ref{energyeq}) and (\ref{pressureeq}) to find
expansions of the energy density $\rho=\rho^{(0)} +\frac{\rho^{(3)}}{a^3}+
\frac{\rho^{(6)}}{a^6}+\cdots$ and similarly for pressure
$p$. (ii) Identify  each term in the energy density and pressure with perfect fluid with barotropic equation of state
and impose $p^{(3n)}=(n-1)\rho^{(3n)}~ (n=0,1,2,\cdots)$. This gives  results of $h^{(3n)}-f^{(3n)}$.
(iii) Substitution of these results into (\ref{cosmomax}) yields $f^{(3n)}$ and $h^{(3n)}$ and consequently, energy density and pressure expansions. (Recall we have chosen $A=B$).
In the zeroth order, we obtain the same relations as (\ref{solex}). Higher order expressions
are given by ($\xi\neq 1/2$)
\begin{align}
f^{(3)} &= \left({\sqrt{3\vert\alpha\vert\over 2} \over -2\vert\alpha\vert+3\xi^2 f_*^2}+{\sqrt{3\over 8} m_f \over \sqrt{\vert\alpha\vert} f_*}\right){B\over m^2} \,, \\
f^{(6)} &= C_{f}^{(1)}{B\over m^2}f^{(3)} + C_{f}^{(2)} f^{(3)2}, ~~\cdots,
\end{align}
nd
\begin{align}
h^{(3)} &= f^{(3)}
+ \left(\frac{3 \xi^2f_*}{-2 \vert\alpha\vert+3\xi^2f_*^2}\right) {B\over m^2} \,, \\
h^{(6)} &= f^{(6)} -\frac{1}{2} \left(h^{(3)}-f^{(3)}\right) \left(h^{(3)}-3 f^{(3)}\right)\,, ~~\cdots
\end{align}
with
\begin{align}
C_{f}^{(1)} &= {3\xi^2 f_* \over 2\vert\alpha\vert}
\left({2 \vert\alpha\vert+\xi^2 f_*^2 \over -2\vert\alpha\vert+3\xi^2 f_*^2}\right)
\left({
2\sqrt{6}\vert\alpha\vert-3\xi^2f_*\left(8 \sqrt{\vert\alpha\vert }-\sqrt{6} f_*\right)
\over
2\sqrt{6}\vert\alpha\vert+\xi^2f_*\left(8 \sqrt{\vert\alpha\vert }+3\sqrt{6} f_*\right)
}\right) \,, \\
C_{f}^{(2)} &= {3\xi^2 f_*^2 \over\vert \alpha\vert}
\left({
-(2\xi^2+1)\sqrt{6}\vert\alpha\vert+\xi^2f_*\left(2 \sqrt{\vert\alpha\vert }-\sqrt{6}\xi^2f_*\right)
\over
2\sqrt{6}\vert\alpha\vert+\xi^2f_*\left(8 \sqrt{\vert\alpha\vert }+3\sqrt{6} f_*\right)
}\right) \,.
\end{align}
Putting these expressions into Eqs. (\ref{energyeq}) and (\ref{pressureeq}), we obtain the following evolution equations (keeping terms up to $1/a^6$)
\begin{align}
3H^2 &\simeq {\vert\alpha\vert m^2 \over 3 \xi^2}
+\frac{\left(m_f+{4\vert\alpha\vert f_* \over -2\vert\alpha\vert+3\xi^2 f_*^2}\right)B}
{ a^3}
+{\left({2\vert\alpha\vert+3\xi^2 f_*^2\over -2\vert\alpha\vert+3\xi^2 f_*^2}\right) Bf_*f^{(3)}\over a^6} \,,\label{enereq} \\
-3H^2-2\dot{H} &\simeq {-\vert\alpha\vert m^2 \over 3 \xi^2}
+{\left({2\vert\alpha\vert+3\xi^2 f_*^2\over -2\vert\alpha\vert+3\xi^2 f_*^2}\right) Bf_*f^{(3)}\over a^6} \,\label{preeeq}.
\end{align}
We make a couple of comments regarding the above expansions. The first one is that   all of the above expansions including
energy density and pressure
are given in powers of $B/m^2$, $x^{(3n+3)}/x^{(3n)}\sim B/m^2,$
$x^{(3n)}=(f^{(3n)},h^{(3n)}, \rho^{(3n)},p^{(3n)} )$.
This feature persists  for the higher expansion coefficient not included.
Therefore, one can say that the expansions make sense as long as $B/m^2<<1,$
and in this case, contributions from the higher order terms rapidly decay with the expansion of the Universe and can be neglected.
Second one is that the evolution equations mimic $\Lambda$CDM behavior up to $1/a^3$ term{\footnote{Restoring Planck mass, the second term in the right hand side of Eq. (\ref{enereq}) is written as $\rho^{(3)}=
\left(\frac{m_f}{M_p}+{4\vert\alpha\vert f_*/M_p \over -2\vert\alpha\vert+3\xi^2 f_*^2/M_p^2}\right)M_pB.$ For $B/m^2M_p<<1, m_f<< M_p,$ and ${\cal O(\xi), {\cal O}(\vert\alpha\vert)}\sim 1$, that we are interested in, $\rho^{(3)}\sim M_pB= \frac{B}{m^2M_p}m^2M_p^2<<\Lambda M_p^2$ cannot account for the whole of the dark matter density but only a very small fraction of it. Therefore, unknown dark matter sector has to be added separately in this approach and we assume that this is the case from here on.}}.
Therefore, higher order terms starting from $1/a^6$ in Eqs. (\ref{enereq}) and (\ref{preeeq}) can be interpreted as a deviation from $\Lambda$CDM,
and
 if they are very small, it may be possible that such deviation is compatible with observational data. In the next section, we perform data analysis in order to check this explicitly. For that purpose, we only keep expansions up to $1/a^6.$

Before passing, we make the following remark.
The numerators in the previous expansions vanishes for $\xi=1/2$, and this case must be treated separately.
One can check that $A$ also has to be zero and  the expansion coefficients $f^{(3n)},h^{(3n)}, (n=1,2\cdots) $
cannot be determined uniquely, that is, they are arbitrary. This does not define a consistent expansion and this case is discarded.
For general values of  $\xi< 1/2$,  $1/a^6$ term contributes a {\it positive} energy density for $m_f<< M_p$.

\section{\noindent Data Analysis}






In this section we explore whether or not having additional energy density  of the form $\rho_s=p_s\sim1/a^6$  is consistent
with current observations and if it is, what is the permissible density
parameter $\Omega_s$.
We use the recent observational data such as type Ia supernovae (SN),
baryon acoustic oscillation (BAO) based on large-scale structure of galaxies,
cosmic microwave background radiation (CMB), and Hubble constant.
For spatially flat $\Lambda\textrm{CDM}$ model with the additional energy density, the Friedmann equation
is given by
\begin{equation}
   \frac{H^2(z)}{H_0^2}
   = \Omega_{r}(1+z)^4+\Omega_{m}(1+z)^3 + \Omega_{\Lambda} + \Omega_{s} (1+z)^6,
\label{eq:hubble}
\end{equation}
where $z \equiv a_0/a-1$ is the redshift with $a(t)$ the cosmic expansion
scale factor, $H \equiv \dot{a}/a$ is the Hubble parameter with $H_0$
its present value, and $\Omega_{i}$ with $i=r$, $m$, $\Lambda$, $s$
indicates the current density parameter for radiation, matter, cosmological constant,
and the deviation, respectively.
We assume that $\Lambda$ has relaxed to its constant value and include radiation component.
In our model we have four free parameters, which are denoted
as $\bth=(h,\Omega_b,\Omega_\Lambda,\Omega_s)$;
$h$ is a normalized present-day Hubble parameter,
$H_0 = 100h~\textrm{km}~\textrm{s}^{-1}~\textrm{Mpc}^{-1}$; $\Omega_b$
is the baryon density parameter; the density parameter of cold dark matter is
given as $\Omega_c=1-\Omega_b-\Omega_\Lambda-\Omega_s$ and
$\Omega_m=\Omega_b+\Omega_c$.
During exploring the parameter space, we use $\log_{10}\Omega_s$ as the free
parameter for the deviation with a prior $\log_{10}\Omega_s > -25$.
The lower bound is chosen as the limit where models with and without
the deviation cannot be distinguished within the current observational precision.
To obtain the likelihood distributions for model parameters, we use the
Markov chian Monte Carlo (MCMC) method based on Metropolis-Hastings
algorithm to randomly explore the parameter space that is favored by
observational data \cite{MCMC}.
The method needs to make decisions for accepting or rejecting
a randomly chosen chain element via the probability function
$P(\bth|\mathbf{D}) \propto \exp(-\chi^2/2)$,
where $\mathbf{D}$ denotes the data,
and $\chi^2=\chi_\textrm{SN}^2 + \chi_\textrm{BAO}^2 + \chi_\textrm{CMB}^2
+ \chi_{H_0}^2$ is the sum of individual chi-squares
for SN, BAO, CMB, and $H_0$ data (defined below).
During the MCMC analysis, we use a simple diagnostic to test the convergence
of MCMC chain: the means estimated from the first (after buring process)
and the last 10\% of the chain are approximately equal to each other
if the chain has converged (see Appendix B of Ref.\ \cite{Abrahamse-etal-2008}).

For SN data set, the Union2.1 sample with 580 members is used
\cite{Suzuki-etal-2012}.
We use the chi-square that is marginalized over the zero-point uncertainty
due to absolute magnitude and Hubble constant \cite{SNIa-MRG}:
\begin{equation}
    \chi_{\textrm{SN}}^2 = c_1 - c_2^2 / c_3,
\end{equation}
where
\begin{equation}
     c_1 = \sum_{i=1}^{580}
       \left[ \frac{\mu(z_i;\bth)-\mu_\textrm{obs}(z_i)}{\sigma_i}\right]^2,
      \quad
     c_2= \sum_{i=1}^{580}
        \frac{\mu(z_i;\bth)-\mu_\textrm{obs}(z_i)}{\sigma_i^2},
       \quad
     c_3= \sum_{i=1}^{580} \frac{1}{\sigma_i^2},
\end{equation}
$\mu_\textrm{obs}(z_i)$ and $\sigma_i$ the distance modulus and
its error of SN at redshift $z_i$, respectively,
$\mu(z;\bth)=5\log [(1+z) r(z)]$ the model distance modulus, and
$r(z)$ the comoving distance at redshift $z$,
\begin{equation}
   r(z)=\frac{c}{H_0 \sqrt{\Omega_k}}
       \sin \left[\sqrt{\Omega_k}\int_0^z \frac{H_0}{H(z')} dz' \right].
\end{equation}
Here $c$ is the speed of light, and $\Omega_k$ is the current density parameter
for spatial curvature ($\Omega_k=0$ in this paper).

We use an effective distance measure which is related to the BAO scale
\cite{Eisenstein-etal-2005},
\begin{equation}
   D_V(z) \equiv \left[r^2 (z) \frac{cz}{H(z)} \right]^\frac{1}{3}
\end{equation}
and a fitting formula for the redshift of drag epoch ($z_d$)
\cite{Eisenstein-Hu-1998}:
\begin{equation}
   z_d = \frac{1291(\Omega_m h^2)^{0.251}}{1+0.659(\Omega_m h^2)^{0.828}}
       \left[1 + b_1 (\Omega_b h^2)^{b_2} \right],
\end{equation}
where
\begin{equation}
   b_1 =0.313 (\Omega_m h^2)^{-0.419}
       \left[1+0.607(\Omega_m h^2)^{0.674} \right], \quad
   b_2 =0.238 (\Omega_m h^2)^{0.223}.
\end{equation}
As the BAO parameter, we use the six numbers of $r_s(z_d)/D_V(z)$ extracted
from the Six-Degree-Field Galaxy Survey \cite{6dFGS},
the Sloan Digital Sky Survey Data Release 7 and 9 \cite{SDSS},
and the WiggleZ Dark Energy Survey \cite{WiggleZ},
where $r_s(z)$ is the comoving sound horizon size.
These BAO data points were used in the WMAP 9-year analysis
\cite{WMAP9}.
Since the sound speed of baryon fluid coupled with photons ($\gamma$)
is given as
\begin{equation}
   c_s^2 = \frac{\dot{p}}{\dot\rho}
         = \frac{\frac{1}{3}\dot\rho_\gamma}{\dot\rho_\gamma + \dot\rho_b}
         = \frac{1}{3\left[1+(3\Omega_b/4\Omega_\gamma)a\right]},
\end{equation}
the comoving sound horizon size before the last scattering becomes
\begin{equation}
   r_s(z) = \int_0^t c_s dt'/a
          = \frac{1}{\sqrt{3}} \int_0^{1/(1+z)}
               \frac{da}{a^2 H(a)[1+(3\Omega_b/4\Omega_\gamma)a]^\frac{1}{2}},
\end{equation}
where $\Omega_\gamma=2.469\times 10^{-5} h^{-2}$ for the present CMB
temperature $T_\textrm{cmb}=2.725~\textrm{K}$.
The BAO data points together with an inverse covariance matrix between
measurement errors can be written in vector and matrix forms as \cite{WMAP9}
\begin{eqnarray}
    &&\mathbf{d}_\textrm{BAO}=
       \left( \begin{array}{c}
             r_s(z_d)/D_V(z=0.1) \\
             D_V(z=0.35)/r_s(z_d) \\
             D_V(z=0.57)/r_s(z_d) \\
             r_s(z_d)/D_V(z=0.44) \\
             r_s(z_d)/D_V(z=0.60) \\
             r_s(z_d)/D_V(z=0.73) \end{array} \right)
      =\left( \begin{array}{c}
             0.336 \\
             8.88 \\
             13.67 \\
             0.0916 \\
             0.0726 \\
             0.0592 \end{array} \right),  \\
    &&\mathbf{C}_{\textrm{BAO}}^{-1} =
          \left( \begin{array}{cccccc}
          4444.4 & 0 & 0 & 0 & 0 & 0  \\
             0   & 34.602 & 0 & 0 & 0 & 0  \\
             0   & 0 & 20.661157 & 0 & 0 & 0  \\
             0   & 0 & 0 & 24532.1 & -25137.7 & 12099.1  \\
             0   & 0 & 0 & -25137.7 & 134598.4 & -64783.9  \\
             0   & 0 & 0 & 12099.1 & -64783.9 & 128837.6 \end{array} \right).
\end{eqnarray}
The chi-square is given as $\chi_{\textrm{BAO}}^2=\mathbf{d}_{\textrm{BAO}}^T
\mathbf{C}_{\textrm{BAO}}^{-1} \mathbf{d}_{\textrm{BAO}}$.

We also use the CMB distance priors based on WMAP 9-year data for
testing our model (see Ref.\ \cite{WMAP9} for the detailed
description).
The first distance measure is the acoustic scale $l_A$ defined as
\begin{equation}
   l_A = \pi \frac{r(z_*)}{r_s (z_*)}.
\end{equation}
The decoupling epoch $z_*$ can be calculated from the fitting function
\cite{Hu-Sugiyama-1996}:
\begin{equation}
   z_*=1048 [1+0.00124(\Omega_b h^2)^{-0.738}]
           [1+g_1(\Omega_m h^2)^{g_2}]
\end{equation}
where
\begin{equation}
   g_1 = \frac{0.0783(\Omega_b h^2)^{-0.238}}{1+39.5(\Omega_b h^2)^{0.763}},
   \quad
   g_2 = \frac{0.560}{1+21.1(\Omega_b h^2)^{1.81}}.
\end{equation}
The second distance measure is the shift parameter $R$ which is given by
\begin{equation}
   R(z_*) = \frac{\sqrt{\Omega_m H_0^2}}{c} r(z_*).
\end{equation}
According to WMAP 9-year observations \cite{WMAP9},
the estimated values and inverse covariance for the three parameters
($l_A$, $R$, and $z_*$) are given as
\begin{equation}
    \mathbf{d}_\textrm{CMB}=
       \left( \begin{array}{c}
             l_A (z_*) \\
             R(z_*) \\
             z_*    \end{array} \right)
      =\left( \begin{array}{c}
             302.40 \\
             1.7246 \\
             1090.88    \end{array} \right), \quad
    \mathbf{C}_{\textrm{CMB}}^{-1} =
          \left( \begin{array}{ccc}
          3.182   & 18.253 & -1.429 \\
          18.253  & 11887.879 & -193.808 \\
          -1.429  & -193.808 &  4.556 \end{array} \right).
\end{equation}
The chi-square is given as $\chi_{\textrm{CMB}}^2=\mathbf{d}_{\textrm{CMB}}^T
\mathbf{C}_{\textrm{CMB}}^{-1} \mathbf{d}_{\textrm{CMB}}$.

In order to impose a tight constraint on the Hubble constant parameter,
we apply a Gaussian prior on the Hubble constant,
$H_0=73.8\pm 2.4~\textrm{km}~\textrm{s}^{-1}~\textrm{Mpc}^{-1}$
(68\% CL) \cite{Riess-etal-2011}. The chi-square is given as
$\chi_{H_0}^2 = [(H_0-73.8)/2.4]^2$.

\begin{figure*}
\mbox{\epsfig{file=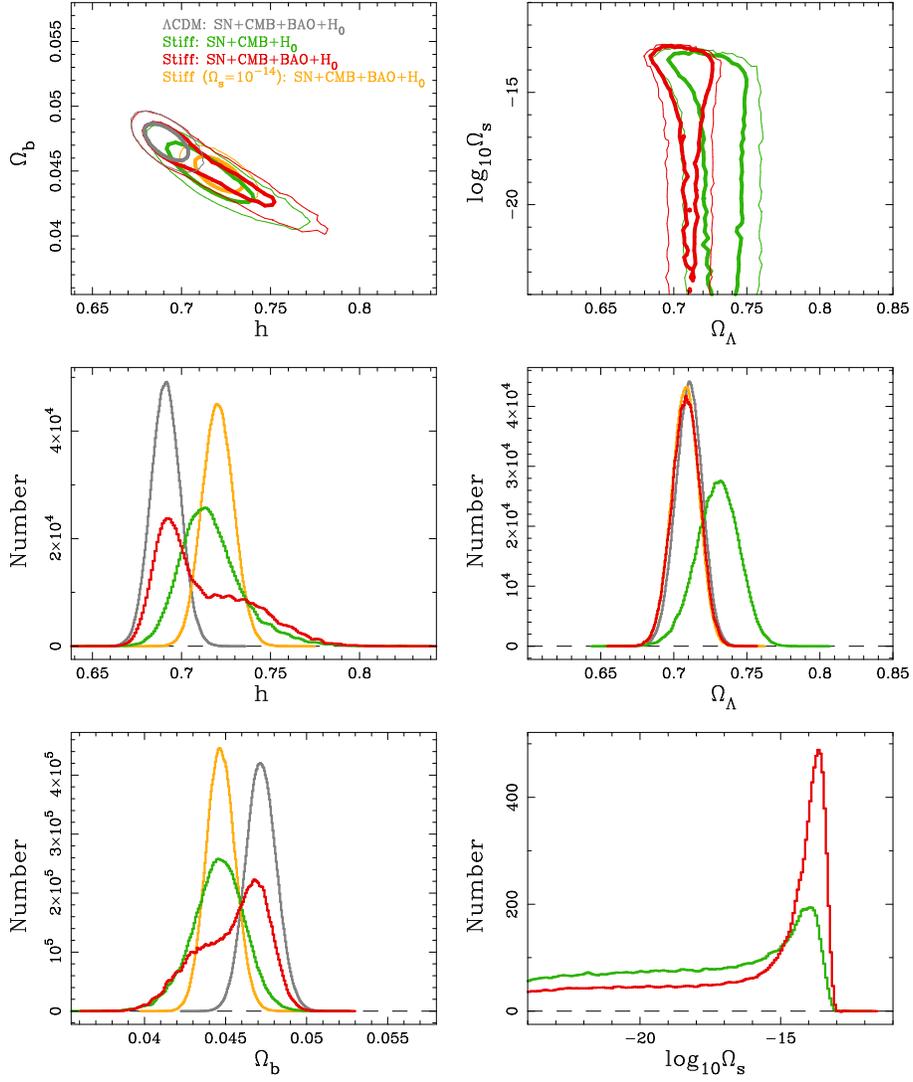,width=120mm,clip=}}
\caption{
         Two-dimensional likelihood contours (top) and one-dimensional
         histograms (middle and bottom panels) of cosmological parameters
         in the flat $\Lambda\textrm{CDM}$ model with the deviation.
         In the two-dimensional likelihood contours, thick and thin curves
         denote the 68.3\% and 95.4\% confidence limits, respectively.
         The green and red color codes indicate the results from the
         combined analysis with SN+CMB+$H_0$ and SN+CMB+BAO+$H_0$, respectively.
         The yellow color represents the likelihoods of model parameters
         with a fixed value of $\Omega_{s}=10^{-14}$, obtained with
         SN+CMB+BAO+$H_0$ data sets.
         The $\Lambda\textrm{CDM}$ result is shown as gray color for comparison.
         Since the dark matter density parameter is simply given
         as $\Omega_c \simeq 1-\Omega_b-\Omega_\Lambda$, the likelihood distributions
         for $\Omega_c$ are omitted.
         }
\label{fig:Stiff-SBCH1235}
\end{figure*}

Figure \ref{fig:Stiff-SBCH1235} shows one- and two-dimensional likelihood
distributions of model parameters in the flat $\Lambda\textrm{CDM}$ model with
the deviation. Using the MCMC method, we constrain our model parameters
$\bth=(h,\Omega_b,\Omega_\Lambda,\Omega_s)$
with the combined data sets. The results are shown as green contours for
SN+CMB+$H_0$ data sets and red for SN+CMB+BAO+$H_0$.
As expected, the joint analysis with all combination of data sets gives
tighter constraints on parameters.
To be consistent with current observations, the amount of the additional energy density
should be small with $\Omega_s \lesssim 10^{-13}$.
In the $h$- and $\Omega_b$-likelihood distributions,
the deviation model (red) appears to favor the parameter regions
similar to those of $\Lambda\textrm{CDM}$ model (gray curves).
However, we note that the model favors regions of larger Hubble constant
and smaller baryon density as the amount of the deviation increases.
Interestingly, the current observations prefer the deviation with positive energy density with
$\Omega_s \sim 10^{-14}$, with a minimum chi-square smaller than
the $\Lambda\textrm{CDM}$ one, and the preference for positive deviation
is amplified by adding BAO data points (see bottom-right panel
of Fig.\ \ref{fig:Stiff-SBCH1235}).
For SN+CMB+BAO+$H_0$ data sets, the best-fit locations in the parameter
space are $\bth=(h,\Omega_b,\Omega_\Lambda,\Omega_s)
=(0.738,0.0433,0.708,2.36\times 10^{-14})$ with
$\chi_\textrm{min}^2=567.96$ (minimum chi-square) for the deviation model,
and $\bth=(0.692,0.0472,0.711,0)$ with $\chi_\textrm{min}^2=572.31$
for $\Lambda\textrm{CDM}$ model, respectively.
We also present the likelihood results for the deviation model
with a fixed value of $\Omega_s = 10^{-14}$ (yellow), which
shows different $h$--$\Omega_b$ likelihood, compared with the case of
freely varying $\Omega_s$ (red contours).
As stated above, fixing the deviation component with
$\log_{10} \Omega_s=-14$ favors Hubble constant (baryonic matter density)
that is slightly larger (smaller) by the amount of about 5\% than
$\Lambda\textrm{CDM}$ value (yellow and gray curves).

Very recently, the CMB data of Planck satellite has been publicly available
together with a new result for cosmological parameter estimation
\cite{Planck-2013}. Since the cosmological parameters used
in this paper are consistent with the Planck results within the observational
precision, we expect that our constraint on the positive deviation will not change much even with the new observational data.
We defer a more detailed analysis using Planck data for future research.

\section{Conclusion and discussions}

In this paper, we investigated cosmology of massive spinor electrodynamics
with torsion which is non-minimally coupled with the vector field.
The massive vector field and torsion cooperate together to
generate dark energy and dark matter. Depending on the values of the vector-torsion coupling constant, the cosmology reveals several novel  aspects which are not present in the massless spinor electrodynamics alone.

The analysis  shows that  the theory effectively provides   dark energy  and pressureless dark matter  with additional energy density\ of the form $\rho_s\sim 1/a^6$.
Comparisons with observations show that its existence is consistent, even though only a very small portion of energy density $\Omega_s\sim 10^{-14}$ is allowed.
If we apply this data analysis  to  (\ref{enereq}), we see that $B/m^{2}\sim 10^{-7}$ for  ${\cal O(\xi), {\cal O}(\vert\alpha\vert)}\sim 1$.
Perhaps, the importance of the deviation lies not in the magnitude of its portion but in the possibility that there could be other type of component in the current Universe.



The restriction $\vert\xi\vert<1/2$ and dependence of cosmology on $\xi$
are hard to  be understood intuitively. But
$\xi$ can be introduced in a different context.  If we consider a different model where the trace of the torsion field $S_\mu$ is replaced with another massive vector field $B_\mu$ along with its field strength, then,  in Eq. (\ref{gafi})  $\xi$ can be interpreted as a coupling constant between the two vectors $A_\mu$ and $B_\mu$.
The ensuing cosmology will show the same qualitative behavior, because the field strength of $B_\mu$ is zero for isotropic cosmology, and  it can duplicate the cosmological behavior of $S_\mu.$
But a new restriction on $\xi$ now depending on the mass of $B_\mu$ will emerge and it would be worthwhile to explore this aspect in detail
in connection with a possible meaning of $\xi$ in cosmology.

In  Eq. (\ref{enereq}), the dark energy contribution comes from the massive vector field and torsion
whereas  dark matter   also contains a contribution from the vector-spinor interaction term. Therefore,
cosmology of  massive spinor electrodynamics effectively provides an interacting dark energy and dark matter model whose interaction is via the
well-known standard vector-spinor interaction.
It would be interesting to compare with other models in the literature \cite{Micheletti:2009pk} and to explore whether massive spinor electrodynamics can be another but a more realistic field theoretical model of interacting dark energy.


We only considered non-minimal torsion coupling with the vector field.
 There exist a couple of other sources of  non-minimal couplings and extensions one could associate with massive spinor electrodynamics. The first one is
to include  direct interactions between the vector field and curvatures
\cite{Hellings:1973zz}. The other is to assume non-minimal couplings of torsion  with fermionic sector \cite{russel} and  non-vanishing of the totally anti-symmetric torsion components. Taking these  into account may open up new possibilities for cosmology.

It remains to be seen whether the cosmology of massive spinor electrodynamics discussed in this paper can provide a viable description of dark energy of current Universe. But if this is possible, then, our analysis predicts   some deviation from $\Lambda$CDM   with additional contributions of the energy density and pressure, $\rho_s=p_s\sim1/a^6$ given in Eqs.
(\ref{enereq}) and (\ref{preeeq}), even though  they occupy  only a very small fraction  of our Universe. But at least, its existence
is favored with a slightly better statistical probability than $\Lambda$CDM as is discussed in Sec. IV.

The last discussion
concerns  the expression of  dark energy in terms of
mass of the vector field given by Eq. (\ref{coss}).
With a choice of negative $\alpha$,
 the  dark energy density which is  responsible for a repulsive force is proportional to  mass squared of the vector field.
This is attributed to an existence of the cosmological solution associated with classical background configuration  $A=(f_*,0,0,0)$,  $S_\mu=(h_*,0,0,0)$, and $\bar\psi=\psi=0,$ in which the specific non-minimal coupling effectively transforms cosmological term of (\ref{cosco}) into (\ref{coss}) with the relation (\ref{hfconstants}) being satisfied.
For  values like ${\cal O}(\vert\alpha\vert)\sim {\cal O}(\xi)\sim 1,$
$m\sim 10^{-61}M_p$   can yield the current dark energy density, and   it is within the allowed range of the {\it photon} mass limit
$~~\mu\lesssim 10^{-62}~\textrm{kg}$ \cite{Goldhaber:2008xy}. It can be made more flexible with adjustments of the parameters $\alpha$ and $\xi,$
 in which case it can encompass the  massive dark photon \cite{Petraki:2014uza}. Therefore, massive (dark) QED can be a realistic model for the dark energy and if this is the case, our result is suggesting that nonvanishing mass of the photon  or dark photon could be responsible for  the dominant component of our Universe.



\section{Acknowledgments}
We would like to thank anonymous referee for valuable suggestions,
 and Young-Hwan Hyun, Joohan Lee, Tae Hoon Lee,  Seokcheon Lee,
Taeyoon Moon, and Wanil Park for useful discussions.
P. O. was supported by the Basic Science Research Program through the
National Research Foundation of Korea (NRF) funded by Ministry of Education,
Science and Technology (2010-0021996) and by a NRF grant funded by the Korean
government (MEST) through the Center for Quantum Spacetime (CQUeST) of Sogang
University with Grant No. 2005-0049409.
C.G.P. was supported by Basic Science Research Program through the National
Research Foundation of Korea (NRF) funded by the Ministry of Science, ICT
and Future Planning (No.\ 2013R1A1A1011107) and was partly supported
by research funds of Chonbuk National University in 2012.


\end{document}